\documentclass[a4paper]{article}

\usepackage{INTERSPEECH2021,url}
\usepackage{graphicx}

\usepackage[backref=false]{hyperref}
\hypersetup{plainpages = false,
              breaklinks=true,
              linktocpage,
              colorlinks   = true, 
              urlcolor     = blue, 
              linkcolor    = blue, 
              citecolor   = red, 
              pdfstartpage=1, 
              pdfstartview=FitH,  
              pdfkeywords={}}

\title{The Phonexia VoxCeleb Speaker Recognition Challenge 2021 System
Description}
\name{Josef Slav\'{i}\v{c}ek, Albert Swart, Michal Kl\v{c}o, Niko Br\"ummer}
\address{
  Phonexia s.r.o, Brno, Czech Republic
  }
\email{josef.slavicek@phonexia.com}
\graphicspath{ {./} }
\begin{document}
\maketitle
\begin{abstract}
We describe the Phonexia submission for the VoxCeleb Speaker Recognition Challenge 2021 (VoxSRC-21) in the self-supervised speaker
verification track. Our solution was very similar to IDLab's winning submission for VoxSRC-20. An embedding extractor was bootstrapped using momentum contrastive learning, with input augmentations as the only source of supervision. This was followed by several iterations of clustering to assign pseudo-speaker labels that were then used for supervised embedding extractor training. Finally, a score fusion was done, by averaging the zt-normalized cosine scores of five different embedding extractors. We briefly also describe unsuccessful solutions involving i-vectors instead of DNN embeddings and PLDA instead of cosine scoring.  
\end{abstract}

\section{Introduction}
Our submission for the VoxSRC-21 self-supervised track\footnote{See track 3 at \url{https://www.robots.ox.ac.uk/~vgg/data/voxceleb/competition2021.html}.} was based on IDLab's submission to VoxSRC-20~\cite{thienpondt2020idlab}. The main differences were:
\begin{itemize}
    \item We did more iterations in stage 2 and we tried various modifications of training setup during stage 2. This eventually led to a training setup similar to stage 3 from \cite{thienpondt2020idlab} that we used for some of the last training iterations in stage 2. So we have no distinct stage 3. 
    \item We used score normalization to improve the accuracy of the baseline cosine scoring backend. Cohorts were populated by random selection from the training data. 
\end{itemize} 
Details follow in several sections later. Table~\ref{table:result} gives comparison of our VoxSRC-21 result against the other two top scoring submissions. A comparison is also given with the result of the best team in the supervised track, as well as the supervised and self-supervised tracks of VoxSRC-20.\footnote{\url{https://www.robots.ox.ac.uk/~vgg/data/voxceleb/competition2020.html}}

\subsection{Things that didn't work}
Our first attempt was to use i-vectors with cosine scoring~\cite{dehak09_interspeech}, because this provides a speaker verifier that can be trained without speaker labels. However, with this technique, we could not get any result below 20\% EER. Apart from being a less flexible model than modern DNN embedding extractors, this solution also does not benefit from augmentation---it does not use labels and therefore cannot use the fact that different augmentations of the same input must be of the same speaker.      

Later, when we had already trained a DNN embedding extractor that gave reasonably accurate results using cosine scoring, we tried to train a PLDA scoring backend~\cite{Kenny_HTPLDA_2010,Brummer_Odyssey_2010}, with a number of different self-supervised training techniques. These techniques included: 
\begin{itemize}
    \item using pseudo-speaker labels obtained via clustering (k-means and spectral clustering)
    \item variational Bayesian PLDA~\cite{JonasPLDA}
    \item an approximate EM algorithm for PLDA training that marginalizes over the hidden labels in the E-step, by using a Gibbs sampler, to sample from: 
    $$P(\text{labels}\mid \text{PLDA model}, \text{unsupervised embeddings})$$ 
\end{itemize}
We got no benefit from any of these techniques, relative to plain cosine scoring. The only unsupervised technique that did work to improve cosine scoring was score normalization, as will be explained below. 

\begin{table}[ht!]
\centering
\begin{tabular}{|l|c|c|}
        \hline
        VoxSRC-21 track 3 & EER [\%] & minDCF\\
        \hline
        Team 1 & \textbf{5.6} & 0.34 \\
        \emph{Phonexia} & 6.5 & \textbf{0.32} \\
        Team 3 & 6.9  & 0.37 \\
        \hline
        \hline
        2021 track 1 & 1.8 & 0.10\\
        2020 track 1 & 3.7 & 0.18\\
        2020 track 3 & 7.2 & 0.34\\
        \hline
\end{tabular}
\caption{Best three VoxSRC-21 self-supervised results, vs best result in tracks 1 (supervised) and 3 (self-supervised) of 2021 and 2020.}
\label{table:result}
\end{table}

\section{Training description -- overview}
Self-supervised training of the embedding extractor was done in two stages. Stage 1 (also referred to as iteration 0) uses contrastive learning, where input augmentation provides the only form of supervision. Stage 2 is iterative, over several iterations. Each iteration takes the embeddings generated by the previous iteration, then clusters them into a number of pseudo-speakers (we used 7500 clusters). These pseudo-speaker labels are then used to train the next embedding extractor, using a supervised training algorithm. Details follow below.

\subsection{Data}
As required by the rules, we used the VoxCeleb 2 training set,\footnote{\url{https://www.robots.ox.ac.uk/~vgg/data/voxceleb/vox2.html}} \emph{without} speaker labels, for training all system parameters and for populating the score normalization cohorts. This training set has roughly one million speech segments, spoken by almost 6000 speakers. Although the number of speakers is known, we did not use this as a parameter to our system. 

We used the labelled, official VoxSRC-21 validation set to monitor minDCF and EER and to make various decisions about which algorithms and system variants to use. The validation set contains 60~000 verification trials, about half of which are targets and the rest non-targets.

To better understand the utility of \emph{equal-error-rate} (EER) for assessing speaker verification accuracy, see~\cite{EERpaper}.

\subsection{Stage 1 -- contrastive learning}

At the level of general description, there is no difference between our setup and IDLab's solution from the last year. So we refer the reader to subsection 2.1.1 of~\cite{thienpondt2020idlab}. In contrastive learning, multiple different embeddings are manufactured from every training speech segment, by using a variety of different input augmentations. Positive scores are formed by cosine scoring of two embeddings (with different augmentations) from the same input. Negative scores are formed by cosine scoring of two (augmented) embeddings from different inputs. The training objective is a multiclass cross-entropy for the pretext task of finding a positive score amongst a very large number of negative scores~\cite{moco}.     

\subsection{Stage 2 -- iterative clustering}

We follow the approach described in subsection 2.1.2 of~\cite{thienpondt2020idlab} with some modifications : stage 2 consists of several iterations where each iteration has two steps. In the first step, a neural network from the previous iteration (or from stage 1 in the case of the first iteration) is used to generate speaker embeddings for train utterances. By clustering these embeddings using k-means, followed by agglomerative hierarchical clustering (AHC), we created speaker pseudo-labels. Like IDLab, we fixed the number of pseudo-speakers to 7500. In the second step we used these pseudo-labels to train discriminatively the neural network as speaker classifier. Then we iterate, using this neural network as embedding extractor for next iteration.

One difference between our approach and IDLab's is that we are training two neural networks in parallel. In each iteration we are using speaker pseudo-labels from network A as targets for the next iteration of network B and vice versa. This approach is inspired by \cite{dividemix}.

The second difference is that we used more iterations in stage 2 than IDLab. When we observed that the pace of improvements between iterations slows down, we introduced some ad-hoc changes of training setup which, we assumed, help the convergence. Unfortunately, there was no time to verify if these changes actually improved the results beyond the level of statistical noise. In the end we obtained a better final system in stage 2 than last year's IDLab's system, but it was at the cost of very complicated pipeline where many steps might be unnecessary or even detrimental.

\section{Fusion}
A number of different systems were trained as explained above, by varying architectures and training procedure details. A selection of these systems was eventually combined for the final submission, using score fusion. The selection of systems to fuse was made by scoring on the official VoxSRC-21 development set. All systems were scored using cosine scoring. Adaptive ZT-norm \cite{matejka2017scorenorm} score normalization\footnote{We found ZT-norm gave slightly better results than s-norm.} was applied to each individual system using a cohort of 10000 utterances randomly selected from the training dataset. Mean and standard deviation statistics for each normalization step were calculated after discarding the top 10 cohort scores and using the top 200 of the remaining scores.

System fusions were created by averaging the ZT-normalized scores of the individual systems. Table \ref{table:results} lists the EERs of the final submission as well as the systems that made it to this fusion. The individual systems are described in more detail in Section \ref{section:stage2}.

\section{Training description -- implementation}
Further details of our DNN embedding extractor training follow below.

\subsection{Training data processing}

As input features we used 80-dimensional filterbanks computed from 25ms frames with 10ms step. We applied mean normalization with a floating window of 300 frames, followed by global mean \& variance normalisation.

We used two different augmentation setups. The first processed utterances on-the-fly and selected one of following options with equal probabilities for each utterance: 
\begin{itemize}
    \item leave the utterance unchanged,
    \item reverberate it using impulse response from RIRS \cite{rirs},
    \item add noise from MUSAN-noise \cite{musan2015} with SNR values randomly chosen from $\lbrace 10, 5\rbrace$ , 
    \item add noise from MUSAN-music with SNR values chosen randomly from $\lbrace 10, 7, 5 \rbrace$. 
\end{itemize}

In the second augmentation setup, augmented copies of training data were prepared in advance. We created 3 additional copies from the clean training data using the following augmentations:
\begin{itemize}
    \item reverberation using the impulse responses from RIRS \cite{rirs}
    \item added noise from MUSAN-noise and MUSAN-music \cite{musan2015} with SNR values randomly chosen from $\lbrace 10, 7, 5 \rbrace$
    \item audio coder/decoder application using FFmpeg compression libraries.
\end{itemize}

Most of our training scripts use 400 frame chunks as training examples. For training example preparation, we used our implementation similar to the KALDI-like egs generation.  We define a ``training epoch'' as roughly 1.66 million utterances. All the utterances were sampled uniformly from the training data.

\subsection{Stage 1 -- contrastive learning}

For stage 1, we used the first, on-the-fly augmentation setup and we ran the training for one epoch (1.66 million utterances).

The neural network architecture was ResNet50 \cite{resnet} with the modification described in the JHU paper~\cite{magneto}, where the numbers of the output filters in the individual ResNet stages are 128, 128, 256, and 256. We used Snyder's pooling layer~\cite{snyderxvectors} with the modification of the standard deviation calculation (Eq.\eqref{eq:std}) that helps us to make the pooling layer numerically stable by clamping the input of square root, where $\epsilon=\textrm{1.0e-3}$:
\begin{equation}
    \sigma = \sqrt{\max(E[x^2] - E[x]^2, \epsilon)}
    \label{eq:std}
\end{equation}
The dimensionality of the speaker embedding was 256.

The system was trained on 6 GPUs, with batch size 8 per GPU, and without batch normalization synchronization among GPUs. The nominal learning rate (LR) per GPU was 0.0125 and we used the following schedule: during the first 10\% of training, the LR grows linearly from 0 to the nominal value. During the next 23.33\% of training it remains constant. The remaining time is split into 10 steps and between steps it decays to a half of the previous value. The loss function used in this stage is equation (2) from \cite{thienpondt2020idlab} with scale $s=10$. The size of the MoCo~\cite{moco} queue was 65536 and the MoCo momentum was 0.999. The system was trained in mixed precision (\texttt{torch.cuda.amp}) with the SGD optimizer, with Nesterov momentum 0.9 and weight decay 0.0001. We used shuffling batch normalization~\cite{moco}.

\subsection{Stage 2 -- iterative clustering}
\label{section:stage2}

Our stage 2 roughly follows technique established in \cite{thienpondt2020idlab}, sec 2.1.2 and 2.2.2. We used the same clustering algorithm implementation (k-means, followed by AHC) and the same number of output clusters (7500). 

For the most of our stage 2 iterations, we trained two neural networks separately in each iteration. \emph{Network A} was trained, using the first, on-the-fly, augmentation setup, while \emph{network B} was trained using the second augmentation setup with pre-augmented copies of the training data. In each iteration we used pseudo-labels generated from network A as the targets for network B and vice versa. Each network was initialized with the weights from the previous iteration as described in \cite{thienpondt2020idlab}, the last paragraph of 2.2.2.

For stage 2, we used ResNet34 with the JHU modifications (numbers of output channels were 128, 128, 256, 256), with the standard Snyder's pooling layer.\footnote{Here, instead of the clamping in~\eqref{eq:std}, $\epsilon=\textrm{1.0e-4}$ was instead added to the input of the square root.} The embedding dimensionality was still set at 256. It was trained on four GPUs with batch size 32 per GPU and without batch normalization synchronization. The learning rate was 0.0125 per GPU, and the LR scheduler the same as in stage 1. The network was trained in mixed precision (\texttt{torch.cuda.amp}) with the SGD optimizer with Nesterov momentum 0.9 and weight decay 0.0001. 

The loss function, used in this stage, was a combination of BiTempered loss~\cite{bitemperedloss} and Angular Additive Margin Softmax (AAM-Softmax) \cite{aamsoftmax}. From the latter we adopt the formula for logits computation:
\begin{align}
\begin{split}
    z &= s (\cos\theta - m), \;\;\textrm{for targets}  \\
    z &= s (\cos\theta), \;\;\;\;\;\;\;\;\;\,\textrm{for non-targets} 
\end{split}
\end{align}
where $\cos\theta$ is the cosine distance  between an embedding and a speaker representative in the classification head. The cosine sore was scaled by $s=\textrm{40}$ and the margin was $m=\textrm{0.2}$. In our framework, calculation of the cosine distance is implemented as the dot-product between length-normalized embeddings. To make it numerically stable, we calculate the norm (length) of the embedding, $\mathbf x$, as:
\begin{equation}
    |\mathbf x| = \sqrt{\sum_{k=1}^{n}{|x_k|^2} + \epsilon}
    \label{norm}
\end{equation}
where $\epsilon=\textrm{1.0e-4}$. The logits, $z$, serve as input to the BiTempered loss~\cite{bitemperedloss}, with parameters $t_1=\textrm{0.9}$ and $t_2=\textrm{1.1}$.
This was our basic setup for the stage 2 and we used it for the first 5 iterations. In iterations 6 -- 8 we changed the learning rate, specifically $\text{LR}_\text{nominal}=\textrm{0.125}$ per GPU.

From iteration 9 we used another setup, which moved the recipe slightly towards the setup used in IDLab's stage 3. We switched from BiTempered loss to AAM-Softmax as described in~\cite{aamsoftmax}, with bigger margin $m=\textrm{0.3}$. From iteration 9 onwards, we no longer initialized weights between iterations and the LR was set to 0.0125 per GPU.

From iteration 11 onwards, we modified our pseudo-label generation. As the input to clustering we used concatenated embeddings from our two networks. We still obtained two different sets of labels from clustering the training dataset because of the random initialization of k-means.

\begin{figure}[ht!]
    \centering
    \includegraphics[scale=0.25]{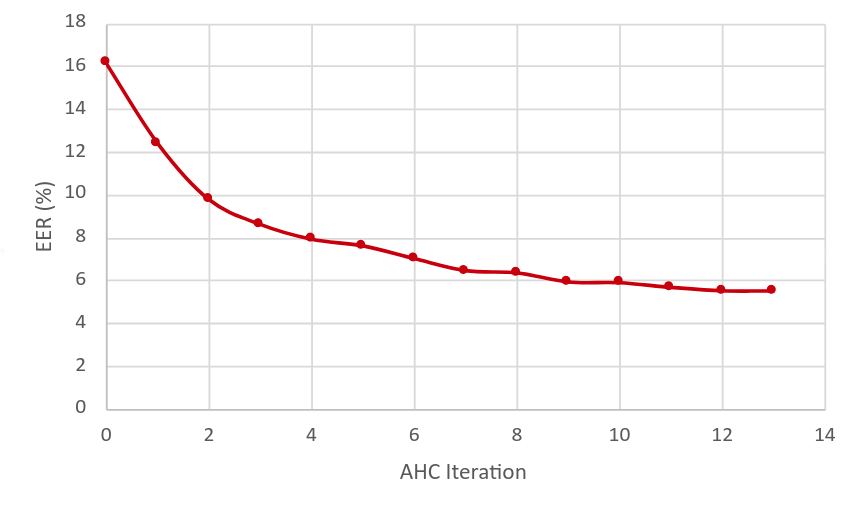}
    \caption{EER of iterative clustering on the VoxSRC-21
validation set.}
    \label{fig:iter}
\end{figure}

In iteration 13, we used 6 seconds long training examples instead of 4 seconds. Figure \ref{fig:iter} and Table \ref{table:iter} show the performance of networks in each iteration.


\section{Final fusion and competition result}
After iteration 13 our team was still in 4th position (ranked by EER) in the competition. To climb up on the leaderboard, we trained several systems with minor modifications and fused them. The effect of the individual modifications were very small, almost negligible, but together they let us move from 3rd position to 2nd. On the leaderboard, the relative difference in EER between the 2nd and 3rd teams\footnote{This entry for the 3rd team was eventually withdrawn. Our table~\ref{table:result} shows the final leaderboard result at the close of the competition.} was 0.0012.\\

\noindent The systems we used in final fusion were:
\begin{itemize}
    \item \texttt{iter11}: System after iteration 11 
    \item \texttt{iter13A}: System after iteration 13 
    \item \texttt{iter13B}: Another system after iteration 13 (the difference between \texttt{iter13A} and \texttt{iter13B} is different random initialisation and different augmentation) 
    \item \texttt{iter14}: One more iteration of training, we switched back to shorter training chunks (400 frames) and we tried to downweight utterances which clusterings with different random initialisation put into non-corresponding clusters. 
    \item \texttt{iter15bt}: Final iteration. This time we initialised our network with weights from \texttt{iter14}, we used BiTempered loss with $t_1=\textrm{0.99}, t_2=\textrm{1.01}$ and we used a snapshot of the NN in the state after 60\% of training. 
\end{itemize}

\begin{table}[ht!]
    \centering
    \begin{tabular}{|l|c|c|}
        \hline
        System & EER [\%] & EER after zt-norm [\%]\\
        \hline
        \texttt{iter11} & 5.619 & 5.155 \\
        \texttt{iter13A} & 5.519 & 5.085 \\
        \texttt{iter13B} & 5.468 & 5.089 \\
        \texttt{iter14} & 5.498 & 5.058 \\
        \texttt{iter15bt} & 5.319 & 4.885 \\
        \hline
        Final fusion & 4.986 & 4.508 \\
        \hline
    \end{tabular}
    \caption{Performance of the individual systems and the final fusion -- EER on the VoxSRC-21
validation set.}
    \label{table:results}
\end{table}

\begin{table}[!ht]
    \centering
    \begin{tabular}{|r|r|r|l|}
        \hline
        Iter & EER [\%] & $\Delta$ [\%] & Modifications \\
        \hline
        0 & 16.123 & - & \\
        1 & 12.108 & 23,1 & \\
        2 & 9.887 & 21.2 & \\
        3 & 8,816 & 11.9 & \\
        4 & 8,080 & 8.1 & \\
        5 & 7,645 & 3.9 & \\
        6 & 7,252 & 7.8 & LR 0.125 \\
        7 & 6,624 & 8.0 & LR 0.125 \\
        8 & 6,587 & 1.6 & LR 0.125 \\
        9 & 6,190 & 6.7 & AAM-Softmax, $m=\textrm{0.3}$, \\
        & & & from scratch, \\
        10 & 6,164 & 0.6 & AAM-Softmax, $m=\textrm{0.3}$, \\
        & & & from scratch, \\
        11 & 5,822 & 3.7 & AAM-Softmax, $m=\textrm{0.3}$,\\
        & & & from scratch, \\
        & & & concat. embdeddings \\
        12 & 5,770 & 2.8 & AAM-Softmax, $m=\textrm{0.3}$,\\
        & & & from scratch, \\
        & & & concat. embdeddings \\
        13 & 5,882 & 0.2 & AAM-Softmax, $m=\textrm{0.3}$,\\ 
        & & & from scratch, \\
        & & & concat. embdedds, 6s train egs \\ 
        \hline
    \end{tabular}
    \caption{EER and relative improvement of EER $\Delta$ on VoxSRC-21 validation set in stage 2. The modifications column summarizes changes of the training setup with respect to our basic training setup.}
    \label{table:iter}
\end{table}

\section{Tabula rasa learning vs adaptation}
We conclude with a comment on the self-supervised task. The tabula rasa,\footnote{clean slate} self-supervised training of a speaker verifier is impressive, scientifically interesting and indeed almost magic. If it could be made to work on an even larger scale, it could help to train speaker recognition systems with orders of magnitude more data than is possible on supervised data sets.

There are however some caveats. First, it should be noted that the whole exercise is most probably undoable without the help of a \emph{labelled} validation set. Second, although the labels from the training data have been removed, the data is still carefully hand-curated---we know it has many examples of almost all of the speakers.

Finally, although there will always be new challenging speech modalities---where obtaining speaker labels is hard or expensive---we \emph{do} have lots of labelled out-of-domain data from which to adapt to new domains. In the real world, tabula rasa learning is not enforced for this problem. We would like to propose instead (or in addition) an \emph{adaptation} task, which makes use of the following data resources:
\begin{itemize}
\item A large, \emph{labelled}, out-of-domain training set.
\item Three in-domain data sets: 
\begin{itemize} 
\item Medium-to-large-sized, \emph{unlabelled}, adaptation data. 
\item Small, \emph{labelled} validation set. This set can be used to monitor progress, but also to learn a few system parameters. 
\item Evaluation data.
\end{itemize}
\end{itemize}
In 2013, in the i-vector era, there was a JHU CLSP workshop\footnote{\url{https://www.clsp.jhu.edu/workshops/13-workshop/speaker-and-language-recognition}} that included a domain adaptation challenge for speaker recognition. Maybe it is worthwhile revisiting this problem with the new data sets and algorithms that are available today.

\bibliographystyle{IEEEtran}

\bibliography{mybib}

\end{document}